\documentclass[pra,letterpaper,onecolumn]{revtex4-1}   
\usepackage{amsbsy,amssymb,amsmath,bm,bbold} 
\usepackage{graphicx,color,epsfig,rotate} 
\usepackage{fancyhdr} 
\usepackage{epstopdf}
\usepackage{csquotes}
\usepackage{float}
\usepackage[colorlinks=true,linkcolor=blue,citecolor=blue,urlcolor=blue]{hyperref}
\usepackage{amsmath}

\begin{document}

	\title{Exotic light dynamics around a fourth order exceptional point}
	
	\author{Sibnath Dey, Arnab Laha and Somnath Ghosh}
	\email{somiit@rediffmail.com}
	\affiliation{\vspace{0.3cm} Unconventional Photonics Laboratory, Department of Physics, Indian Institute of Technology Jodhpur, Rajasthan-342037, India
		}

\begin{abstract}
The physics of exceptional point (EP) singularities, has been a key to a wide range of unique physical applications in open systems. In this context, the mutual interactions among four coupled states around a fourth-order EP (EP4) in a physical system is yet to be explored. Here, we investigate the unique features of an EP4 in a fabrication feasible planar optical waveguide with a multilayer gain-loss profile based on only two tunable parameters. A unique `fourth-order $\beta$-switching' phenomenon due to quasi-static gain-loss variation around EP4 has been explored. An exclusive chiral light dynamics following the dynamical variation of the gain-loss profile has been reported for the first time, which enables a special type of asymmetric higher-order mode conversion scheme. Here, all the coupled modes associated with an EP4 are fully converted into different specific higher-order modes based on the choice of encirclement directions. The proposed scheme would present EP4 as a new light manipulation tool for integrated photonic devices.    
\end{abstract}
	
	\maketitle

\section{Introduction}

Nonconservative or open systems behold productive physical significance as they interact with the environment and hence are dissipative or active. Non-Hermitian quantum mechanics describes such an open physical system by converting it into an effective Hamiltonian \cite{book1,book2}. In a particular system, the non-Hermiticity can be introduced by integrating gain-loss. Recently, the attractive features of the non-Hermitian quantum analogy have fertilized the platform for the topological study of various quantum-inspired photonic devices. One of the astonishing topological features of such type of photonic systems is the appearance of branch point singularities namely, exceptional points (EPs)~\cite{Heiss_Avoided,Heiss_repulsion_of_resonance} in the system parameter space. An EP of order $n$ can be defined as a particular branch point singularity in system's parameter space, where $n$ number of eigenvalues and the corresponding eigenstates simultaneously coalesce and hence the effective Hamiltonian of the underlying system becomes defective \cite{book1,book2,Heiss_Avoided,Heiss_repulsion_of_resonance}. However, it has been established that the functionality of an EP of order $n$ can be realized with simultaneous presence $(n-1)$ second-order EPs \cite{Mular_N_EP,Heiss_2008,EP3_sayan}. In this paper, we conventionally use the abbreviation `EP' to define a second-order exceptional point, whereas the higher-order exceptional points are abbreviated as `EP$n$' with $n=3,4$ (i.e., for third- and fourth-order EP). Thus, an EP can be considered as a special type of degeneracy which is different from conventional Hermitian degeneracy called diabolic point (DP), where only the coupled eigenvalues coincide. A DP and an EP are different due to the self-adjointness of the Hamiltonian in the former and the lack of it in the later case~\cite{phase_wave_Heiss}.

The exotic physical phenomena associated with EPs have extensively been explored in a wide range of open system, including optical waveguide \cite{Ghosh16,Doppler16,optical_wg_Chen2015,optical_wg_PRX,my_nonlinearity,Gandhi:20}, photonic crystals~\cite{PC_slabs}, lasers~\cite{laser_PRL,laser_APL_kominis,laser2_PRL}, microcavities~\cite{stadian_shape_microcavity,microcavity_arnab,EP3_microcavity}, and also in several non optical systems, such as atomic~\cite{atomic_spectra} and molecular spectra~\cite{molecular_spectra}, and microwave systems \cite{PhysRevE.69.056216,microcavity_experiment_PRL}. However, conventional photonic systems with sophisticated fabrication control can provide a highly promising and productive domain to study the fundamental aspects of EPs with several unique technological applications such as, asymmetric mode conversion/switching~\cite{Doppler16,Ghosh16,optical_wg_PRX,Doppler16,optical_wg_Chen2015,my_nonlinearity,nonlinear_PRR}, dark-state lasing~\cite{dark_state_lasing}, ultra sensitive EP aided sensing~\cite{sensor1,sensor2,sensor3_appli,sensors_gen_theory}, nonreciprocity enhancement~\cite{Thomas16,optical_non_reciprocity,isolator}. Instead of only second-order EPs, an immense theoretical effort have been put forward to investigate higher-order EPs~\cite{EP3_Graefe,atomic_spectra,EP3_sayan,EP4_sayan}. In this context, third-order EPs (EP3s) attracts attention due to its cube-root response, which is more sensitive to an external perturbation in comparison to a square-root response near a second-order EP \cite{sensor1,sensor2}. The presence of EP3s have recently been reported in microcavity \cite{EP3_microcavity} and waveguide \cite{schnabel} systems.

The non-trivial topological behavior alongside different order of single or multiple EPs can be explored by an adiabatic encirclement process with a quasi-static variation of coupling parameters along a closed loop. Such an EP encirclement scheme results in the adiabatic permutation among the corresponding coupled states where they successively exchange their own identities~\cite{Laha:17,micro_laha,EP3_sayan,EP4_sayan,atomic_spectra,Ghosh16,Gandhi:20,Laha18}. During such an EP-aided state-flipping, one of the eigenstates acquire an additional $\pm\pi$ phase at the end of the encirclement process \cite{phase_wave_Heiss,Atom_cavity_prl}. However, when we consider time or analogous length scale dependent (dynamical) parametric variation around an EP, the adiabaticity of the system's dynamics breaks down. Here the competition of the dynamical EP-encirclement process with the adiabatic theorem leads to an asymmetric mode conversion phenomenon where only one eigenstate follows the adiabatic expectation \cite{gilary_time_dynamics}. An asymmetric mode conversion enabled by a dynamical EP encirclement process refers the conversion of light into a specific dominating mode, regardless of the choice of inputs. Here, the clockwise (CW) and counter-clockwise (CCW) dynamical parametric variation around an EP leads to different dominating modes, beyond the adiabatic restriction \cite{gilary_time_dynamics}. The effect of dynamical parametric encirclement process around a single EP \cite{Ghosh16,Laha18,Gandhi:20,nonlinear_PRR,Doppler16}. or two connecting EPs (or an analogous EP3) \cite{my_nonlinearity,Zhang19} have widely been investigated using various planar or coupled guide-wave systems.     

In this context, the presence of fourth-order EPs (EP4s) deals with more complex topology of the underlying system. It should be quite interesting, if one can manipulate the simultaneous interaction among the four states in the vicinity of an EP4. In general, dynamical encirclements of second- or third- order EPs have attracted enormous attention to develop mode-conversion devices \cite{Ghosh16,Laha18,Gandhi:20,nonlinear_PRR,Doppler16,my_nonlinearity,Zhang19}. Thus, a general question comes out in the context of EP4, i.e., what is the outcome of dynamical encirclement around an EP4? However, with proper parametric control in a planar geometry, the simultaneous interaction among four coupled states and successive state exchange among them around an EP4 has never been explored. Moreover, the device implementation of higher-order mode conversion scheme, using higher-order EPs is still a challenge. 

In this paper, to address the issues mentioned above, we investigate a specially designed gain-loss assisted multi-mode supported planar optical waveguide to host an EP4. Here, non-Hermiticity has been introduced in terms of a multilayer gain-loss profile based on only two control parameters. This is beyond the general prediction for encountering a higher-order EP~\cite{Heiss_2008}, where it was shown that  $(n^2+n-2)/2$ parameters are required to encounter an EP of order $n$. With proper variation of two control parameters, an EP4 has been embedded by encountering three connecting EPs among four coupled modes. We present an unique `fourth-order $\beta$-switching' phenomenon by considering a quasi-static parametric variation around the identified EP4. For the first time, we investigate the unconventional propagation characteristics of the coupled modes following a dynamical parametric encirclement process, enclosing the identified EP4. Here, we explore a special type of chiral dynamics of the coupled modes around the EP4, where four coupled modes (associated with the respective EP4) are converted into a particular dominating higher-order mode, irrespective of the choice of inputs, however, depending on the encirclement direction. An analytical treatment has also been presented to establish the unique state-dynamics around an EP4. The proposed scheme opens up a potential platform to develop higher-order optical mode converters for the integrated or on-chip photonic circuits.  

\section{Analytical Structure of an EP4}

To develop an analogous analytical model to study the appearance of an EP4, we can consider a simple $4\times4$ non-Hermitian Hamiltonian matrix $\mathcal{H}$ having the form $H_0+\lambda  H_q$
\begin{equation}
\mathcal{H}=\left(\begin{array}{cccc}{\epsilon}_1 & 0 & 0 & 0 \\0 &  {\epsilon}_2 & 0 & 0 \\0 & 0 &  {\epsilon}_3 & 0\\0 & 0 & 0 &  {\epsilon}_4\end {array}\right)+\lambda\left(\begin{array}{cccc}0 & \omega_i  & 0 & \omega_j \\ \omega_i & 0 & \omega_k & 0 \\ 0 & \omega_k  & 0 & \omega_l\\ \omega_j & 0 & \omega_l & 0\end {array}\right).
\label{equation_H}
\end{equation}

Here, $H_0$ corresponds to a passive Hamiltonian, which is subjected to a parameter-dependent complex perturbation $H_q$. $\lambda$ represents complex parameter. Here, $H_0$ consist of four passive eigenstates $\epsilon_l (l=1,2,3,4)$. The $H_q$ consist of four interconnected perturbation parameters $w_i, w_j, w_k$ and $w_l$.  Now, four eigenvalues $E_l\,(l=1,2,3,4)$ of $\mathcal{H}$ are calculated by solving the eigen-value equation $|\mathcal{H}-EI|=0$ ($I\rightarrow\,4\times4$ identity matrix) which gives the quartic secular equation
\begin{equation}
E^4+s_1E^3+s_2E^2+s_3E+s_4=0,
\label{equation_E}
\end{equation}
with
\begin{subequations}
	\begin{align}
	s_1&=-\left( {\epsilon}_1+ {\epsilon}_2+ {\epsilon}_3+ {\epsilon}_4\right),
	\label{equation_p1}\\
	s_2&={\epsilon}_1 {\epsilon}_2+ {\epsilon}_2 {\epsilon}_3+ {\epsilon}_3 {\epsilon}_4+ {\epsilon}_4 {\epsilon}_1+ {\epsilon}_1 {\epsilon}_3+ {\epsilon}_2 {\epsilon}_4-\lambda^2\left( \omega_i^2+ \omega_j^2+\omega_k^2+\omega_l^2\right),\label{equation_p2}\\
	s_3&=-\left( {\epsilon}_1 {\epsilon}_2 {\epsilon}_3+ {\epsilon}_2 {\epsilon}_3 {\epsilon}_4+ {\epsilon}_1 {\epsilon}_3 {\epsilon}_4+ {\epsilon}_1 {\epsilon}_2 {\epsilon}_4\right)+\lambda^2\left\{( {\epsilon}_1+ {\epsilon}_2)\omega_l^2+( {\epsilon}_2+ {\epsilon}_3) \omega_j^2+( {\epsilon}_3+ {\epsilon}_4) \omega_i^2+( {\epsilon}_4+ {\epsilon}_1)\omega_k^2\right\},\label{equation_p3}\\
	s_4&={\epsilon}_1 {\epsilon}_2 {\epsilon}_3 {\epsilon}_4-\lambda^2\left( {\epsilon}_1 {\epsilon}_2\omega_l^2+ {\epsilon}_2 {\epsilon}_3 \omega_j^2+ {\epsilon}_3 {\epsilon}_4 \omega_i^2+ {\epsilon}_4 {\epsilon}_1\omega_k^2\right)-\lambda^4\left( \omega_i\omega_l+ \omega_j\omega_k\right)^2.
	\label{equation_p4}
	\end{align}	
\end{subequations}
Using Ferrari's method~\cite{book3}, the roots of the Eq. \ref{equation_E} can be written as
\begin{subequations}\begin{align}
&E_{1,2}=-\frac{s_1}{4}-\eta\pm\frac{1}{2}\sqrt{-4\eta^2-2m_1+\frac{m_2}{\eta}},\label{equation_E12}
\\
&E_{3,4}=-\frac{s_1}{4}+\eta\pm\frac{1}{2}\sqrt{-4\eta^2-2m_1-\frac{m_2}{\eta}};\label{equation_E34}
\end{align}
\label{equation_E1234}
\end{subequations}
where,
\begin{subequations}\begin{align}
&\eta= \frac{1}{2}\sqrt{-\frac{2m_1}{3} +\frac{1}{3}\left(k+\frac{m_3}{k}\right)},
\\
&k=\left(\frac{m_4}{2}+\frac{1}{2}\sqrt{{m_4}^2-4{m^3_3}}\right)^{1/3}.
\end{align}
\end{subequations}
Here, 
\begin{subequations}
\begin{align}
&m_1=-\frac{3s_1^2}{8}+s_2,
\label{equation_m1}\\
&m_2=\frac{s_1^3}{8}-\frac{s_1s_2}{2}+s_4,
\label{equation_m2}\\
&m_3=s_2^2-3(s_1s_3+4s_4),
\label{equation_m3}\\
&m_4=2s_2^3-9s_2(s_1s_3+8s_4)+27(s_1^2s_4+s_3^2).
\label{equation_m4}
\end{align}
\end{subequations}
Eq. \ref{equation_E1234} represents the eigenvalues of $\mathcal{H}$. Now, among the four coupled eigenvalues $E_l\,(l=1,2,3,4)$, we can individually manipulate the interaction among any three chosen pairs such as $\{E_1,E_2\}$, $\{E_2,E_3\}$ and $\{E_4,E_1\}$. Here, three different EPs are identified by the individual coalescence of $E_1$ with $E_2$ or $E_4$, and $E_2$ with $E_3$, by specific settings of perturbation. Under these specific conditions, the system hosts three different EPs through which the eigenvalues (given by Eq. \ref{equation_E1234}) are analytically connected. As it has already been established that the functionality of an EP of order $N$ can be achieved by $(N-1)$ connecting EPs of the order 2 \cite{Mular_N_EP}, the identified three connecting EPs for the chosen pairs $\{E_1,E_2\}$, $\{E_2,E_3\}$ and $\{E_4,E_1\}$ give an analogous effect of an EP4 between $E_l\,(l=1,2,3,4)$. Such situations entitle the validity of the condition 
\begin{equation}
k+\frac{m_3}{k}=0
\end{equation}
 
In the following section, we have implemented the above analytical treatment using the framework of a prototype of a few-mode supported planar optical waveguide with a multilayer gain-loss profile. Here, four chosen modes to encounter an EP4 in the presence of three connecting EPs are analogous to the eigenvalues of $\mathcal{H}$, where the gain-loss profile can be considered as perturbation which controls the coupling among the chosen modes.

\section{Design of a gain-loss assisted Planar Optical waveguide and Encounter of multiple EPs}

\begin{figure}[t]
		\includegraphics[width=8.5cm]{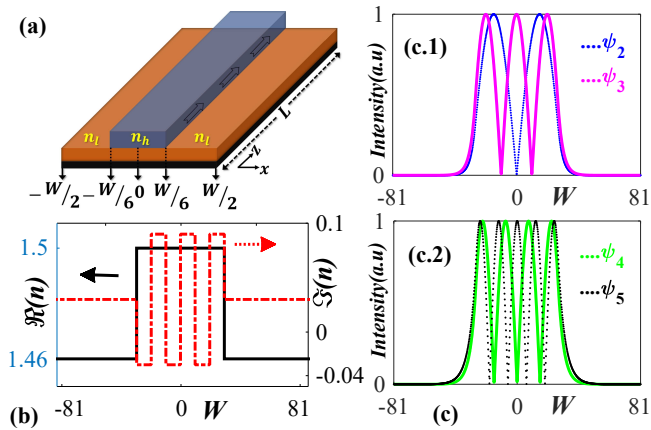}
		\caption{\textbf{(a)} Schematic diagram of the proposed waveguide, having width $W$ (along the transverse axis $x$) and length $L$ (along the propagation axis $z$). $n_h$ and $n_l$ are the passive refractive indices of core and cladding, respectively. \textbf{(b)} The profiles of $\Re(n)$ (solid black line; corresponding to left vertical axis) and $\Im(n)$ (dotted red line; corresponding to right vertical axis) for a specific $\gamma=0.007$ and $\tau=3.2$. \textbf{(c)} Field intensity profiles (normalized) of four chosen modes $\psi_i\,(i=2,3,4,5)$, where (c.1) shows $\psi_2$ and $\psi_3$ and (c.2) shows $\psi_4$ and $\psi_5$.}
		\label{fig1}
\end{figure}

We design a special type of planar optical waveguide, schematically shown in Fig. \ref{fig1}(a), with $x$-axis as the transverse direction and $z$-axis as the propagation direction. The designed waveguide of width $W$ covers the region $-{W/2}\leq x \leq {W/2}$, which consists of a core and a cladding, having the refractive indices $n_h$ and $n_l$ ($n_h>n_l$), respectively. During the operation, we normalize the operating frequency $\omega=1$ and set the total width $W=81\lambda/\pi=162$ (with $\lambda$ as a free space wavelength) and length $L=15\times10^3$ in a dimensionless unit. But conventionally, we choose the micrometer unit for a real prototype.  The passive refractive indices for core and cladding are chosen as $n_h=1.5$ and $n_l=1.46$, respectively. Such a prototype can suitably be realized by thin film deposition technique of glass materials over a thick silica glass substrate. 

Here, the non-Hermiticity has been introduced by a specific transverse distribution of a multilayer gain-loss profile. Inhomogeneity in the gain-loss profile has been controlled by considering two tunable parameters such as $\gamma$ to represent the gain coefficient, and $\tau$ to maintain a fixed loss-to-gain ratio along the transverse direction. The spatially distributed imaginary part of the refractive index profile represents such gain-loss distribution. The multilayer gain-loss profile has been chosen in such a way that the transverse distribution of the complex refractive index profile $n(x)$ can be written as follows:  
\begin{equation}
n(x)=\left\{ 
\begin{array}{ll}\vspace{0.2cm}
n_l+i\gamma,\quad&\textnormal{for}\,\,\,W/6\le |x|\le W/2\\
n_h-i\gamma,\quad&\textnormal{for}\,\left\{ 
\begin{array}{l}-W/4\le x\le -W/6\\-W/8\le x\le 0\\W/8\le x\le W/6 \end{array}\right.\vspace{0.2cm}\\
n_h+i\tau\gamma,\quad&\textnormal{for}\,\left\{ 
\begin{array}{l} -W/6\le x\le -W/8\\0\le x\le W/8\\W/6\le x\le W/4 \end{array}\right.
\end{array}
\right.
\label{nx} 
\end{equation}
i.e., inside the core, $(-{W/6}\le x \le {W/6})$, there are six alternate layers of gain-loss of equal width, whereas the cladding $({W/6}\le|x|\le {W/2})$ consists of only loss.
	
The overall transverse distribution of $n(x)$ has been shown in Fig. \ref{fig1}(b), where solid black line (corresponding to left $y$-axis) and dotted red line (corresponding to right $y$-axis) represents the profiles of $\Re(n)$ and $\Im(n)$ (for a specific $\gamma=0.007$ and $\tau=3.2$), respectively. Each cross-section of the proposed waveguide is associated with a fixed gain-loss distribution that is characterized by a specific set of $\{\gamma,\,\tau\}$, however, the coupling parameters $\{\gamma,\,\tau\}$ can be varied along the $z$-direction to modulate the gain-loss profile for controlling the interactions. Here, the causality condition is satisfied, i.e., the independent tunability of $\Im(n)$ along the propagation axis irrespective of the choice of $\Re(n)$ is realized only at the single operating frequency, as per Kramers-Kronig relation~\cite{kramer_kroning,my_nonlinearity}.  
We have chosen operating parameter in such a way that the waveguide support six scalar modes (linearly polarized) $\psi_j\,(j=1,2,3,4,5,6)$ in increasing order. We compute the propagation constants of the respective modes $\beta_j (j=1\text{---}6)$ by solving the scalar mode equation. According to instantaneous mode profile $\psi(x)$, the scalar mode equation is given by
\begin{equation}
[\partial_x^2+n^2(x)\omega^2-\beta^2]\psi(x)=0
\label{sclar_modal} 
\end{equation}
We choose the small index difference between core and cladding to consider scalar wave approximation, and derive Eq. \ref{sclar_modal} from the Maxwell's equations. Even though the waveguide supports six quasi-guided modes, to encounter an EP4 in the presence of three connecting EPs, we have chosen only four modes $\psi_i\,(i=2,3,4,5)$. Thus, our proposed scalable waveguide system offers the opportunity to encounter multiple EP4s with different combinations of four modes from six quasi-guided modes along with appropriate settings of the gain-loss profile. Input field intensities (normalized) of four chosen modes $\psi_i\,(i=2,3,4,5)$ have been shown in Fig. \ref{fig1}(c). For proper visualization they have been shown in two different plots, where  Fig. \ref{fig1}(c.1) shows $\psi_2$ and $\psi_3$, and in Fig. \ref{fig1}(c.2) displays $\psi_4$ and $\psi_5$.

Here, an EP4 has been embedded by encountering three connecting EPs among four chosen quasi-guided modes. To encounter an EP between two quasi-guided modes, we have identified the abrupt transition between two topologically dissimilar avoided crossings between the corresponding $\beta$-values with crossing/ anticrossing of their real and imaginary prats, i.e., $\Re[\beta]$ and $\Im[\beta]$ for two different $\tau$-values, while varying $\gamma$ within a specified limit. Here, we have to identify two such nearby $\tau$, for which the coupled $\beta$-values exhibit a crossing in $\Re[\beta]$ and a simultaneous anticrossing in $\Im[\beta]$ for particular $\tau$, and the vice-versa, i.e., an anticrossing in $\Re[\beta]$ and a simultaneous crossing in $\Im[\beta]$ for another $\tau$. This is the standard technique to identify an EP \cite{Heiss_Avoided,Ghosh16,Gandhi:20,Laha18}, where for an intermediate $\tau$, two coupled $\beta$-values coalesce. 

Using this standard method, we have identified three connecting EPs among $\psi_i\,(i=2,3,4,5)$. With proper investigations of crossing/anticrossing of $\Re[\beta]$ and $\Im[\beta]$ for two distinct nearby $\tau$-values, we have identified an intermediate $\tau=3.295$ for $\psi_3$ and $\psi_4$, where we observed that the corresponding $\beta$-values coalesce at $\gamma=0.0024$. This indicate the appearance of an EP at $\sim(0.0024, 3.295)$ (say, EP$^{(1)}$) in the $(\gamma,\,\tau)$-plane. In a similar way, we find two other EPs in the $(\gamma,\,\tau)$-plane, such as EP$^{(2)}$ at $\sim(0.0043, 5.002)$ between $\psi_2$ and $\psi_5$, and EP$^{(3)}$ at $\sim(0.012, 2.901)$ between $\psi_2$ and $\psi_3$.    
\begin{figure*}[t]
		\includegraphics[width=16cm]{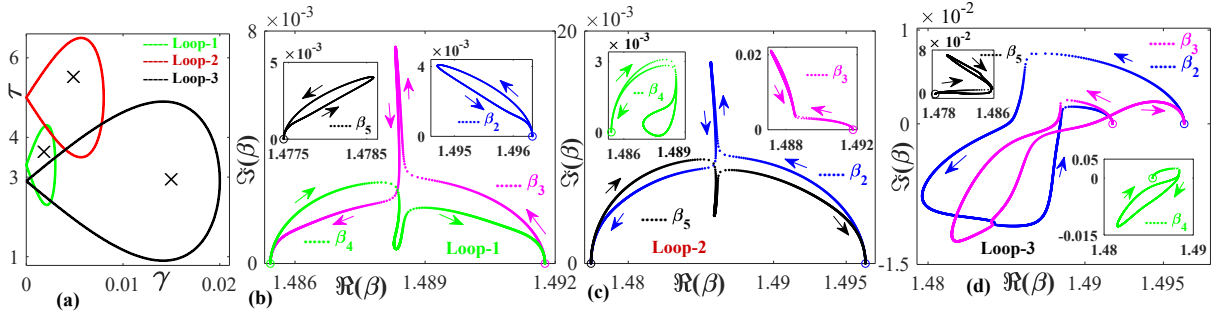}
		\caption{\textbf{(a)} Three chosen parametric loops in $(\gamma,\tau)$-plane to encircle three identified EPs, individually, where Loop-1, Loop-2, and Loop-3 encircle EP$^{(1)}$, EP$^{(2)}$ and EP$^{(3)}$, respectively. Trajectories of complex $\beta_i\,(i=2,3,4,5)$ (represented by dotted blue, pink, green and black curves) in the complex $\beta$-plane following the quasi-static variation of $\gamma$ and $\tau$ along \textbf{(b)} Loop-1, showing the adiabatic switching between $\beta_3$ and $\beta_4$, unaffecting $\beta_2$ and $\beta_5$ (shown in the insets); \textbf{(c)} Loop-2, showing the adiabatic switching between $\beta_2$ and $\beta_5$, unaffecting $\beta_3$ and $\beta_4$ (shown in the insets); and \textbf{(d)} Loop-3, showing the adiabatic switching between $\beta_2$ and $\beta_3$, unaffecting $\beta_4$ and $\beta_5$ (shown in the insets). The circular markers of respective colors represent their initial positions (i.e., for $\phi=0$). Arrows of respective colors indicate their direction of progressions.} 
		\label{fig3} 
	\end{figure*}
	
\section{Modal Propagation characteristics through the waveguide in the presence of multiple EPs}

\subsection{Adiabatic Modal Dynamics: $\beta$-switching}

Here, we study the trajectories of the $\beta$-values of the chosen modes in the complex $\beta$-plane due to the effect of various quasi-static parametric encirclement processes around the identified EPs in the $(\gamma,\tau)$-plane. The shape of the parameter space has been chosen in such a way, so that we consider $\gamma=0$ at both beginning and the end of the encirclement process, which is a crucial requirement for device implementations \cite{gilary_time_dynamics,Ghosh16,Laha18}. To ensure such a condition, we choose a specific contour to enclose single or multiple EPs as
\begin{equation}
\gamma(\phi)=\gamma_{0}\sin\left(\phi/2\right);\,\,\tau(\phi)=\tau_{0}+r\sin\left(\phi\right).
\label{encirclement} 
\end{equation}
where $\gamma_0$, $\tau_0$ and $r$ are three characteristics parameters to enable the stroboscopic variation of $\gamma$ and $\tau$ over the tunable angle $\phi$ ($0\le\phi\le2\pi$). To encircle a particular EP, we have to chose a $\gamma_{0}$ which must be greater than the $\gamma$-value associated with the respective EP. This type of encirclement method is undoubtedly necessary to check the second-order branch point behavior of the identified EPs by scanning the alongside regions around them. The chosen parameter spaces in $(\gamma, \tau)$-plane for encircling each of the identified EPs individually has been shown in Fig. \ref{fig3}(a). Here, Loop-1 (with $\gamma_0=0.003$, $\tau_0 =3.3$, and $r=1$; shown by green contour), Loop-2 ($\gamma_0=0.008$, $\tau_0=5$, and $r=1.5$; shown by red contour), and Loop-3 ($\gamma_0=0.02$, $\tau_0=2.9$, and $r=3$; shown by black contour) individually enclose EP$^{(1)}$, EP$^{(2)}$ and EP$^{(3)}$, respectively.

In Figs. \ref{fig3}(b--d), we have shown the trajectories of $\beta_i\,(i=2,3,4,5)$ following a sufficiently slow variation of $\gamma$ and $\tau$ along the Loop-1. The trajectories of $\beta_i\,(i=2,3,4,5)$ are represented by dotted blue, pink, green and black curves, respectively, where the circular markers of respective colors represent their initial positions at the beginning of the encirclement process (i.e., for $\phi=0$). Each point on the trajectories of $\beta_i\,(i=2,3,4,5)$ are generated by each point evolution of $\gamma$ and $\tau$ along the operating loop. Arrows of respective colors indicate their direction of progressions. 

Now, for a total $2\pi$ complete rotation along Loop-1 (that encloses only EP$^{(1)}$), $\beta_3$ and $\beta_4$ adiabatically exchange their initial positions and form a complete loop in complex $\beta$-plane, as shown in Fig. \ref{fig3}(b), as only $\beta_3$ and $\beta_4$ are analytically connected through EP$^{(1)}$. Such a permutation between two connected $\beta$-values can be called as a $\beta$-switching (analogous to EP-aided flip-of-sates phenomenon) phenomenon. It can be also observed that, $\beta_3$ and $\beta_4$ regain their initial positions for a further one-round encirclement along the same contour. Such a $\beta$-switching phenomenon reveals the second-order branch point behavior of EP$^{(1)}$. Now, the encirclement process following Loop-1 does not affect the trajectories  of $\beta_2$ and $\beta_5$ and hence they remains in the same states making individual loops in the complex $\beta$-plane, as can be seen in two insets of Fig. \ref{fig3}(b). Similar trajectories of $\beta_i\,(i=2,3,4,5)$ have been shown in Fig. \ref{fig3}(c), while varying the $\gamma$ and $\tau$ along Loop-2, which encloses only EP$^{(2)}$. Here, the adiabatic $\beta$-switching has been observed between $\beta_2$ and $\beta_5$ at the end of the encirclement process, as they are analytically connected through EP$^{(2)}$. However, in this case, $\beta_3$ and $\beta_4$ remains in the same states, which have been shown in the insets of the Fig. \ref{fig3}(c). A similar adiabatic $\beta$-switching between $\beta_2$ and $\beta_3$ (two connecting $\beta$'

s through EP$^{(3)}$), unaffecting $\beta_4$ and $\beta_5$ (as shown in the insets), has been exhibited in Fig. \ref{fig3}(d), when we consider the stroboscopic variation of $\gamma$ and $\tau$ along Loop-3 (that encloses only EP$^{(3)}$). In Figs. \ref{fig3}(b--d), the trajectories of unaffecting states for the respective cases have been shown in insets for clear visualization of the trajectories.
\begin{figure*}[t]
		\includegraphics[width=16cm]{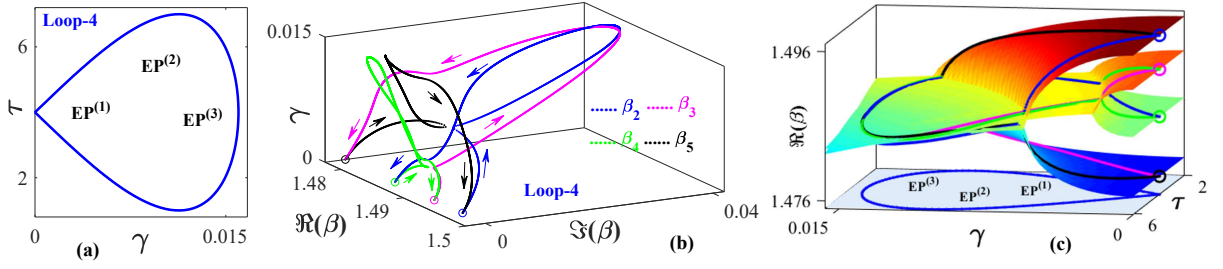}
		\caption{\textbf{(a)} A parametric loop in $(\gamma,\tau)$-plane to encircle three identified EPs, simultaneously. \textbf{(b)} Corresponding trajectories $\beta_i (i=2,3,4,5)$, showing a fourth-order $\beta$-switching phenomenon following the manner $\beta_2 \rightarrow \beta_4 \rightarrow \beta_3 \rightarrow \beta_5 \rightarrow \beta_2$. The colors and markers carry the same meaning as described in the caption of Fig. \ref{fig3}. \textbf{(c)} Topological structure of the Riemann surfaces associated with real parts of $\beta_i\,(i=2,3,4,5)$ as a function of $\gamma$ and $\tau$ within specified ranges. The curves on the surface represent the successive switching of $\beta$-values from their respective surfaces. In the ground surface, the chosen parameter space has been shown separately for proper understanding.}
		\label{fig4}
	\end{figure*}
	
Now, we consider a new parametric loop (say, Loop-4) to encircle all the identified EPs, as shown in Fig. \ref{fig4}(a), for which we have chosen $\gamma=0.015$, $\tau_0=4$, and $r=3$. The associated trajectories of $\beta_i\,(i=2,3,4,5)$ has been shown in Fig. \ref{fig4}(b). Here, with the quasi-static variation of $\gamma$ and $\tau$ along Loop-4, all four propagation constants $\beta_2$, $\beta_3$, $\beta_4$, and $\beta_5$ exchange their own identities and switch successively following the manner $\beta_2 \rightarrow \beta_4 \rightarrow \beta_3 \rightarrow \beta_5 \rightarrow \beta_2$ to make a complete loop in the complex $\beta$-plane. Thus, interestingly the system shows a fourth-order branch point behavior for the corresponding eigenvalues if all three connecting EPs are quasi-statically encircled in the parameter space. This unique type of successive $\beta$-switching phenomenon around three connecting EPs indeed confirms the appearance of an EP4 in system parameter space, where all four chosen modes are analytically connected. Specifically, the successive switching among four $\beta$-values, enabling by an EP4 with the simultaneous presence of three EPs, can be refereed as `fourth-order $\beta$-switching'. Here, we have also observed that if we reverse the encirclement direction (i.e., CCW), then the chosen complex $\beta$-values switch successively following the manner $\beta_2 \rightarrow \beta_5 \rightarrow \beta_3 \rightarrow \beta_4 \rightarrow \beta_2$ (exactly the reverse progressions from the previous observation) in the complex $\beta$-plane. Such type of adiabatic fourth-order $\beta$-switching phenomenon has been reported for the first time in a guided wave system.  
  
In Fig. \ref{fig4}(c), we have demonstrated the formation of the Riemann sheets associated with $\beta_i(i=2,3,4,5)$ by choosing a specific range for both $\gamma$ and $\tau$ (which is sufficient to accommodate Loop-4, and to consider the interaction regime). Here, the overall distribution of $\Re(\beta)$ as a function of $\gamma$ and $\tau$, reveals the simultaneous interaction of $\beta_i(i=2,3,4,5)$ through three connecting EPs. The variation of $\Re(\beta)$ associated with the fourth-order $\beta$-switching phenomenon, as shown in Fig. \ref{fig4}(b), has been shown (by the dotted blue, pink, green and black curves) on the associated distribution of the Riemann surfaces, where it is evident that the real parts of $\beta_i(i=2,3,4,5)$ switch successively following the manner $\beta_2 \rightarrow \beta_4 \rightarrow \beta_3 \rightarrow \beta_5 \rightarrow \beta_2$ from their respective surfaces. For a clear understanding, the parameter space (including the location of three EPs) has been been shown separately in the ground surface of Fig. \ref{fig4}(c) [exactly same loop shown in Fig. \ref{fig4}(a)].         

\subsection{Dynamical parametric encirclement: beam propagation dynamics}

The $\beta$-switching phenomenon due to quasi-static encirclement of the coupling parameters around single or multiple EPs, as described in the preceding section, follow the adiabatic theorem. However, if we consider the time dependence (dynamical) in the parametric variation, then the breakdown of inversion symmetry in the overall gain-loss variation along the time scale compete with adiabatic theorem \cite{gilary_time_dynamics}. The breakdown in the adiabatic theorem has been studied around second- and third-order EPs, which enables chiral or nonchiral dynamics \cite{Ghosh16,Laha18,Doppler16,my_nonlinearity,Zhang19}. However, the chiral aspect due to breakdown in system's adiabaticity following a dynamically encircled EP4 has never been reported. In this section, we explore the effect of such dynamical encirclement around the embedded EP4 and study the chiral aspect of the proposed waveguide device for the first time.
\begin{figure}[b!]
	\includegraphics[width=8.5cm]{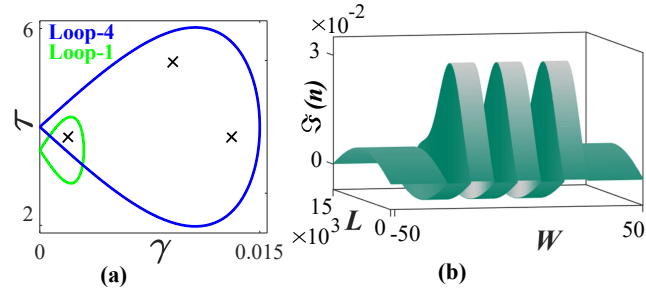}
	\caption{Two chosen parameter spaces in the $(\gamma,\,\tau)$-plane; where Loop-1 encircles only EP$^{(1)}$, and Loop-4 encircles all the identified EPs, i.e., EP$^{(1)}$, EP$^{(2)}$, and EP$^{(3)}$, simultaneously. \textbf{(b)} Length dependent distribution $\Im[n(x,z)]$ after mapping Loop-4.}
	\label{fig5}
\end{figure}

Now, to study the actual propagation of modes through the waveguide, we have to consider a length dependent gain-loss distribution to encircle single or multiple EPs dynamically. Such dynamical encirclement can be achieved by mapping the overall $\Im(n)$ associated with a EP-encircling parameter space throughout the entire length of the waveguide. Here, the propagation of the eigenmodes are governed by the time-dependent Schrodinger equation(TDSE). The parameter space mapping demands the simultaneous closed variation of $\gamma$ and $\tau$ along the $z$-axis throughout the total waveguide length. However, for each transverse cross section (i.e., for a fixed $z$), the refractive index profile corresponds to a specific set of $\{\gamma,\,\tau\}$. Such a parameter space mapping has been achieved by the consideration of $\phi=2\pi z/L$ in Eq. \ref{encirclement} as 
\begin{equation}
\gamma(z)=\gamma_{0}\sin\left[\frac{\pi z}{L}\right];\,\,\tau(z)=\tau_{0}+r\sin\left[\frac{2\pi z}{L}\right].
\label{device1} 
\end{equation}
Here, $Z=0$ and $L$ are associated with $\phi=0$ and $2\pi$, respectively. Thus, one complete encirclement ($0\le\phi\le2\pi$) following Eq. \ref{encirclement} is equivalent to one complete pass ($0\le z\le L$) of light following Eq. \ref{device1} along the propagation length. Here CW encirclement ($0\le\phi\le2\pi$) has been realized by the propagation of light form $z=0$ to $z=L$, whereas the CCW encirclement ($2\pi\le\phi\le0$) has been realized by the propagation of light in the opposite direction, i.e., form $z=L$ to $Z=0$. As for both $Z=0$ and $L$, $\gamma=0$, we can excite and retrieve the passive modes at both the input and output, which is not achievable using conventional circular parametric loop for EP encirclement \cite{Ghosh16,atomic_spectra,Laha18}.

Now, we perform two different dynamical encirclement processes following Loop-1 and Loop-4, as shown in Fig. \ref{fig5}(a) individually. Fig. \ref{fig5}(b) represents the overall variation $\Im[n(x,z)]$ after mapping Loop-4 (following Eq. \ref{device1}) along the entire length of the waveguide. For sufficiently slow variation of $\Im(n)$ along the $z$ direction, the dynamics of the quasi-guided modes are governed by the (1+1)$D$ scalar beam propagation equation, that can be rewritten under the paraxial approximation as:    
\begin{equation}
    2i\omega \frac{\partial \psi(x,z)}{\partial z}= -\left[\frac{\partial^{2}}{\partial x}+ \Delta n^{2}  (x,z) \omega^{2}\right] \psi(x,z),
    \label{paraxial} 
\end{equation}
where $\Delta n^{2}(x,z)\equiv n^{2}(x,z)-n_l^{2}$. We use split steps Fourier method to solve Eq. \ref{paraxial}~\cite{book4}. 

During the dynamical EP-encirclement process, it has been established that only the mode that evolves with the lower average loss follow the adiabatic expectations governed by the associated $\beta$-trajectories. Here the individual decay rate of a particular mode during the evolution around an EP can be calculated by averaging the losses over the entire contour in the complex $\beta$-plane (governed by corresponding quasi-static EP-encirclement process) as \cite{Laha18}
\begin{equation}
    \zeta_{average}= {\frac{1}{2\pi}} \int_{0}^{2\pi} {\Im(\beta)} d\phi
    \label{average loss}
\end{equation}

\begin{figure}[t]
		\includegraphics[width=8.5cm]{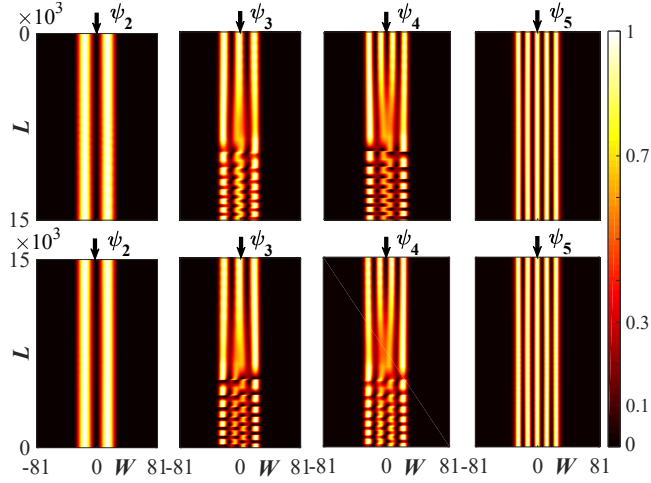}
		\caption{Beam Propagation simulation results for the propagations of $\psi_i\,(i=2,3,4,5)$ following the dynamical parametric encirclement along the Loop-1. The upper panel shows the asymmetric conversions $({\psi_3,\psi_4})\rightarrow\psi_3$, unaffecting $\psi_2$ and $\psi_5$, for propagation from $z=0$ to $z=L$ (i.e., CW dynamical encirclement). The lower panel shows the asymmetric conversions $({\psi_3,\psi_4})\rightarrow\psi_4$, unaffecting $\psi_2$ and $\psi_5$, for propagation from $z=L$ to $z=0$ (i.e., CCW dynamical encirclement). Here, We re-normalize the intensities at each $z$ for visualizing the propagations clearly, and hence the overall effect of loss is not evident.}
		\label{fig6}
\end{figure}

In Fig. \ref{fig6}, we have shown the beam propagation simulation results for the dynamical encirclement process along Loop-1, which encloses only EP$^{(1)}$. Here, to consider the encirclement in the CW direction, we choose input at $Z=0$ to launch the light. The associated beam propagation results of the chosen modes $\psi_i\,(i=2,3,4,5)$ have been shown in the upper panel of Fig. \ref{fig6}. Here, it is evident that, the modes $\psi_3$ and $\psi_4$ associated with EP$^{(1)}$, which are individually excited from $Z=0$ are essentially converted to $\psi_3$ at $z=L$, beyond the adiabatic expectations from the corresponding $\beta$-trajectories shown in Fig. \ref{fig3}(b). Here, as $\beta_4$ evolve with lower average loss (to move to the location of $\beta_3$) in comparison to $\beta_3$, as can be seen in Fig. \ref{fig3}(b) (also calculated by using Eq. \ref{average loss}), the mode $\psi_4$ is adiabatically converted into $\psi_3$, whereas $\psi_3$ shows a nonadiabatic transition (NAT) and is converted into itself. However, the dynamical encirclement along Loop-1 does not affect the propagation of $\psi_2$ and $\psi_5$, and hence they evolve without any conversion. Now for the dynamical encirclement along Loop-1 in the CCW direction (i.e., when the modes are excited from $z=L$), $\psi_3$ evolves adiabatically and is converted to $\psi_4$ (as in this case, $\beta_3$ evolves with the lower average loss in comparison to $\beta_4$), wheres $\psi_4$ behaves nonadiabatically and is converted to $\psi_4$ at $z=0$, as shown in the lower panel of Fig. \ref{fig6}. Here, also $\psi_2$ and $\psi_5$ remains unaffected. Thus, depending on the direction of propagation, we get the conversions $\{\psi_3,\,\psi_4\}\rightarrow\psi_3$ (for $z=0\rightarrow L$; CW rotation) and $\{\psi_3,\,\psi_4\}\rightarrow\psi_4$ (for $z=L\rightarrow 0$; CCW rotation). Thus, the device hosts the chiral property and leads to a higher-order asymmetric mode conversion between two specific higher-order modes. In a similar way, one can observe the asymmetric mode conversion between $\psi_2$ and $\psi_5$ (unaffecting $\psi_3$ and $\psi_4$) by considering the dynamical encirclement along Loop-2, and also between $\psi_2$ and $\psi_3$ (unaffecting $\psi_4$ and $\psi_5$) by considering the dynamical encirclement along Loop-3.
\begin{figure*}[t]
		\includegraphics[width=14cm]{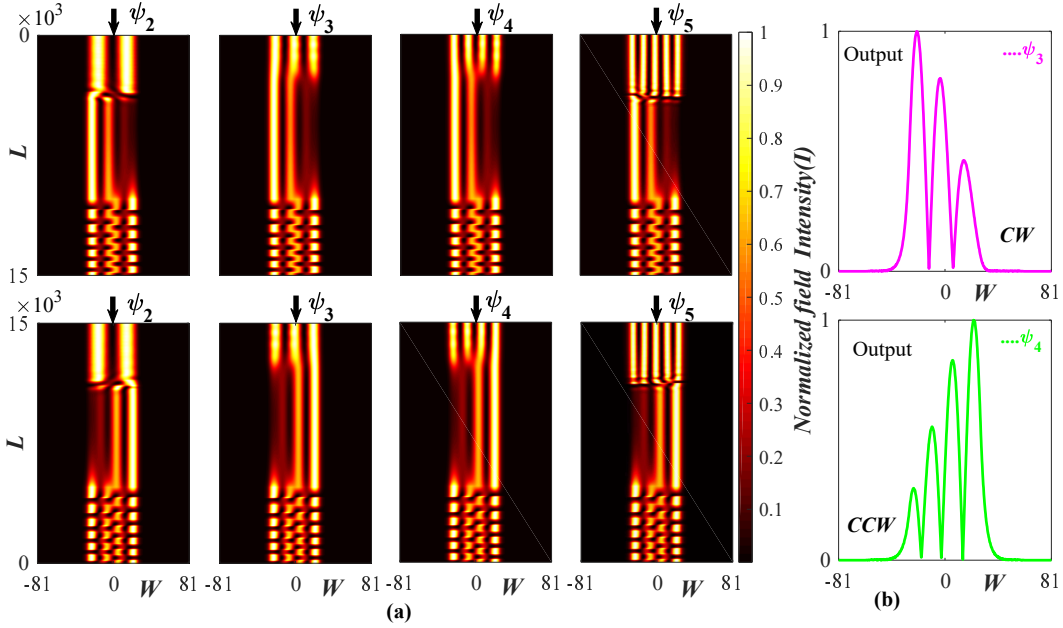}
		\caption{\textbf{(a)} Propagation of four chosen modes $\psi_i(i=2,3,4,5)$ following dynamical EP4-encirclement process along Loop-4. The upper panel shows the asymmetric conversions $\{\psi_2,\psi_3,\psi_4,\psi_5\}\rightarrow\psi_3$, for the encirclement in CW direction (i.e., during propagation from $z=0$ to $z=L$). Whereas, The lower panel shows the asymmetric conversions $\{\psi_2,\psi_3,\psi_4,\psi_5\}\rightarrow\psi_4$, for the encirclement in CCW direction (i.e., during propagation from $z=L$ to $z=0$). \textbf{(b)} The upper panel shows the overall output field intensity at $z=L$ associated with the conversions shown in the upper panel of (a). Whereas, the lower panel shows the overall output field intensity at $z=0$ associated with the conversions shown in the lower panel of (a).}
		\label{fig7}
\end{figure*}

Now, we have shown the effect of dynamical encirclement around the embedded EP4 with the simultaneous presence of three connecting EPs by mapping Loop-4 [as already shown in Fig. \ref{fig5}(b)] along the entire length of the waveguide. The associated beam propagation results and the expected output intensity profiles have been shown in Fig. \ref{fig7}. To consider a CW dynamical encirclement process, the light has been launched at $z=0$, where the propagations of the chosen modes $\psi_i\,(i=2,3,4,5)$ has been shown in the upper panel of Fig. \ref{fig7}(a). Here, it has been shown that all the chosen modes, that are individually excited from $z=0$, are finally converted into dominating $\psi_3$ at $z=L$, i.e., we get the conversions $\{\psi_2,\psi_3,\psi_4,\psi_5\}\rightarrow\psi_3$. The overall output field intensity (normalized) has been shown in the upper panel of Fig. \ref{fig7}(b). Here, only $\psi_4$ behaves adiabatically and is converted to $\psi_3$ and the other modes $\psi_2$, $\psi_3$, and $\psi_5$ follow the NATs and converted into $\psi_3$. Such an adiabatic conversion of $\psi_4\,(\rightarrow\psi_3)$ can also be verified from the associated adiabatic $\beta$-trajectories, shown in Fig. \ref{fig4}(b), where $\beta_4$ evolves with lower average loss (calculated using Eq. \ref{average loss}) in comparison to the others.

If we reverse the encirclement direction by exciting the inputs $\psi_i\,(i=2,3,4,5)$ from $z=L$, then we get the conversions $\{\psi_2,\psi_3,\psi_4,\psi_5\}\rightarrow\psi_4$ after the completion of the propagation (i.e., at $z=0$), beyond the adiabatic expectations governed by the associated $\beta$-trajectories. In this case, we have calculated the average losses of individual modes by using Eq. \ref{average loss}, and observed that $\beta_3$ evolves with lower average loss in comparison to the others. The corresponding beam propagation results have been shown in the lower panel of Fig. \ref{fig7}(a). Here, only $\psi_3$ behaves adiabatically and is converted to $\psi_4$ and the other modes $\psi_2$, $\psi_4$, and $\psi_5$ follow the NATs and converted into $\psi_4$. The overall output field intensity (normalized) for this anticlockwise encirclement process has been shown in the lower panel of Fig. \ref{fig7}(b).   

We also calculate the mode conversion efficiencies in terms of overlap integrals between input and output fields as
\begin{equation}\label{conversion}
  C_{R\rightarrow S}=\frac{\left|\int\psi_R \psi_S dx\right| ^2} {\int\left|\psi_R\right| ^2 dx {\int\left|\psi_S\right| ^2 dx}};\quad\{R,S\} \in i.
\end{equation}
Here, $C_{R\rightarrow S}$ defines the conversion efficiency for the conversion $\psi_R\rightarrow\psi_S$. For the conversions $\{\psi_2,\psi_3,\psi_4,\psi_5\}\rightarrow\psi_3$, as shown in the upper panel of Fig. \ref{fig7}(a), we find the maximum conversion efficiency of 77.18\%. On the other hand, for the conversions $\{\psi_2,\psi_3,\psi_4,\psi_5\}\rightarrow\psi_4$, as shown in the lower panel of Fig. \ref{fig7}(a), we find the maximum conversion efficiency of 86.64\%. 

We also examine the robustness of the such a unique EP4-aided higher-order asymmetric mode conversion process against the parametric tolerances during the fabrication using state-of-the-art techniques. Accordingly, we have appended an average $5\%$ random parametric fluctuations on the dynamical encirclement process along Loop-4 and observed that the higher-order asymmetric mode conversion process, as can be seen in Fig. \ref{fig7}, is omnipresent even in the presence of moderate parametric tolerances.

\section{Analytical treatment For unconventional higher order modal propagation}

Based on only two coupling parameters, the unconventional dynamics of four coupled modes around an EP4 can can analytically be treated as follows. Let assumes that the corresponding four-level Hamiltonian depends on two time-dependent potential parameters $\Gamma_1(t)$ and $\Gamma_2(t)$ (analogous to $\gamma$ and $\tau$). Within the adiabatic limit, the eigenfunctions follow the TDSE during evolutions. Now to estimate the nonadiabatic correction terms during the propagation of four eigenmodes around an EP4, we consider possible conversions among $\psi_R^{ad}(\Gamma_1,\Gamma_2)$ and $\psi_S^{ad}(\Gamma_1,\Gamma_2)$ (with corresponding eigenvalues $\beta_R^{ad}(\Gamma_1,\Gamma_2)$ and $\beta_S^{ad}(\Gamma_1,\Gamma_2)$, respectively), where $\{R,S\}\in\{2,3,4,5\}$. Here, the dynamical nonadiabatic correction terms can be written as 
\begin{subequations}
	\begin{align}
	&\Theta_{R\rightarrow S}^{\text{NA}}=\vartheta_{R\rightarrow S}\exp\left\{+i\oint_0^T\Delta\beta_{R,S}^{\text{ad}}[\Gamma_1(t),\Gamma_2(t)]dt\right\},	\label{Theta1} \\
	&\Theta_{S\rightarrow R}^{\text{NA}}=\vartheta_{S\rightarrow R}\exp\left\{-i\oint_0^T\Delta\beta_{R,S}^{\text{ad}}[\Gamma_1(t),\Gamma_2(t)]dt\right\}; 	\label{Theta2} 
	\end{align}
	\label{theta} 
\end{subequations}
with the pre-exponet terms
\begin{subequations}
	\begin{align}
	&\vartheta_{R\rightarrow S}=\left\langle\psi_R^{\text{ad}}[\Gamma_1,\Gamma_2]\left|\sum_{m=1}^{2}\dot{\Gamma_m}\frac{\partial}{\partial\Gamma_m}\right|\psi_S^{\text{ad}}[\Gamma_1,\Gamma_2]\right\rangle,\label{Theta_v1}\\
	&\vartheta_{S\rightarrow R}=\left\langle\psi_S^{\text{ad}}[\Gamma_1,\Gamma_2]\left|\sum_{m=1}^{2}\dot{\Gamma_m}\frac{\partial}{\partial\Gamma_m}\right|\psi_R^{\text{ad}}[\Gamma_1,\Gamma_2]\right\rangle,\label{Theta_v2}\\
	&\text{and}\,\,\Delta\beta_{R,S}^{\text{ad}}[\Gamma_1,\Gamma_2]=\beta_R^{\text{ad}}[\Gamma_1,\Gamma_2]-\beta_S^{\text{ad}}[\Gamma_1,\Gamma_2]\notag\\
	&\qquad\qquad\equiv\text{Re}[\Delta\beta_{R,S}^{\text{ad}}[\Gamma_1,\Gamma_2]-i\Delta\zeta_{R,S}^{\text{ad}}[\Gamma_1,\Gamma_2]\label{beta}.
	\end{align}
\end{subequations}
In Eq. \ref{theta}, the suffixes $R \rightarrow S$ and $S \rightarrow R$ corresponds to the conversion $\psi_{R}^{\text{ad}} \rightarrow \psi_{S}^{\text{ad}}$ and vice versa, respectively. $T$ is the duration of encirclement. In Eq. \ref{beta}, $\left|\Delta\zeta_{R,S}^{\text{ad}}\right|$ represents the `relative gain' between the two corresponding modes. As the pre-exponent terms of Eq. \ref{theta} contain the time derivative of two potential parameters ($\dot{\Gamma_m};\,m=1,2$) the exponential divergence in $T$ of the exponents of Eq. \ref{theta} beats the $T^{-1}$ suppression associated with the associated pre-exponent terms. Thus, if we consider a situation $\Delta\zeta_{R,S}^{\text{ad}}>0$, then for $T \rightarrow \infty$, $\Theta_{R\rightarrow S}^{\text{NA}} \rightarrow \infty$ and $\Theta_{S\rightarrow R}^{\text{NA}} \rightarrow 0$. Thus, for a very slow parametric encirclement within the adiabatic limit, the conversion associated with $\Theta_{S\rightarrow R}^{\text{NA}}$ follow the adiabatic expectations, whereas the conversion associated with $\Theta_{R\rightarrow S}^{\text{NA}}$ dose not follow the adiabatic expectations. In a similar way, we can establish the vice-versa condition for $\Delta\zeta_{R,S}^{\text{ad}}<0$. Now around an EP, if CW parametric variation gives $\Delta\zeta_{R,S}^{\text{ad}}>0$ then the CCW parametric variation gives $\Delta\zeta_{R,S}^{\text{ad}}>0$. Thus for the dynamical parametric encirclement in any of the directions, only one state evolves adiabatically.

For our proposed waveguide, we can consider $\{R,S\}\in\{2,3,4,5\}$ to calculate the possible nonadiabatic correction terms for different encirclement direction. Now, following the dynamical encirclement along Loop-1 (around only EP$^{(1)}$), we have obtained $\Delta \zeta_{3,4}>0$ for encirclement in CW direction, whereas $\Delta \zeta_{3,4}<0$ for encirclement in CCW direction. Thus for CW dynamical encirclement along Loop-1, the proposed waveguide allows the conversions of $\psi_4\rightarrow\psi_3$ (adiabatic) and $\psi_3\rightarrow\psi_3$ (nonadiabatic). On the other hand, the waveguide allows the adiabatic conversion of $\psi_3\rightarrow\psi_4$ (adiabatic) and $\psi_4\rightarrow\psi_4$ (nonadiabatic) for CCW dynamical encirclement process. As the other two connecting EPs, i.e., EP$^{(2)}$ and EP$^{(3)}$ are away from the chosen parametric loop, the other possible nonadiabatic correction terms can be neglected. The beam propagation results shown in Fig. \ref{fig6} support these analytical predictions.       
Now, we consider the situation of the dynamical encirclement along Loop-4 around the embedded EP4 with the simultaneous presence of three connecting EPs. Here, during the clockwise encirclement we have obtained $\Delta \zeta_{3,4}>0$, which allows the adiabatic conversion of $\psi_4\rightarrow\psi_3$ and nonadiabatic conversion of $\psi_3\rightarrow\psi_3$. However, unlike Loop-1, Loop-4 encircles all the identified EPs, and hence we have to consider the other possible nonadiabatic correction terms. Now, the validity of $\Delta \zeta_{3,4}>0$ demands $\Delta \zeta_{3,2}>0$ and $\Delta \zeta_{3,5}>0$ due to the modification of the decay rates of the coupled modes due to the tailored gain-loss profile in the proposed waveguide. Such additional conditions allows the conversions $\psi_2\rightarrow\psi_3$ and $\psi_5\rightarrow\psi_3$. Thus for encirclement in the clockwise direction, we get the overall conversions $\{\psi_2,\psi_3,\psi_4,\psi_5\}\rightarrow\psi_3$. On the other hand, the CCW encirclement along Loop-4 yields the nonadiabatic correction terms, $\Delta \zeta_{3,4}<0$, in addition with $\Delta \zeta_{4,5}>0$ and $\Delta \zeta_{4,2}>0$. Thus, we get the conversions $\{\psi_2,\psi_3,\psi_4,\psi_5\}\rightarrow\psi_4$, where only $\psi_3$ follows the adiabatic expectation. The beam propagation results shown in the lower panel of Fig. \ref{fig7} support these analytical predictions.   
Thus if we dynamically encircle the embedded EP4 in the simultaneous presence of three connecting EPs, then the device enables a specific chiral light dynamics, where irrespective of the input modes, the coupled modes associated with the EP4 is converted into a specific dominating higher-order mode depending on the direction of light propagation. Here, for the propagation in two different directions, the waveguide delivers two different dominating higher-order mode. Thus the proposed waveguide can be used as an efficient higher-order mode converter, where one can achieve different higher-order modes by changing the direction of light propagations.

\section{Conclusion}
 
In summary, we report an exclusive topologically robust, compact and fabrication feasible scheme for higher-order mode conversion using the framework of a gain-loss assisted and multimode supported planar optical waveguide. Here, introducing a spatial distributing of a multilayer gain-loss profile (based on only two tunable parameters), the interactions among four chosen modes have been manipulated with encounter of three connecting EPs. The fourth-root branch point behavior of the embedded EP4 has been revealed in terms of a successive $\beta$-switching phenomena among four chosen modes following a quasi-static parametric encirclement process around three connecting EPs. Without using any complex topology, the effect of dynamical parametric encirclement around the embedded EP4 has been revealed by customizing only a 2D parameter space associated with the gain-loss profile, which is the first-ever report in the context of a dynamical EP4-encirclement scheme. The waveguide, hosting such a dynamical EP4-encirclement scheme, enables a unique chiral light dynamics, where depending on the direction of light propagation, the associated four coupled modes get converted to a specific dominating higher-order mode, irrespective of the choice of inputs. Here, owing to the device chirality, while operating around an EP4, two different dominating modes survive for propagation in two different direction. Thus, the proposed specially configured waveguide indeed has a vast potential to present itself as a higher-order asymmetric mode converter. Such a customized device is able to excite a particular mode in a multi-modal configuration, that would be suitable for various pumping scheme in integrated (or chip-scale) device applications. The proposed scheme, hosting the rich physics of an EP4, would open up opportunities to boost the device applications associated with next-generation all-optical communication, and computing.  

\section*{Acknowledgments} 
S.D. acknowledges the support from the Ministry of Human Research and Development (MHRD), Government of India. A.L. and S.G. acknowledge the financial support from the Science and Engineering Research Board (SERB) [Grant No. ECR/2017/000491], Department of Science and Technology, Government of India.

\bibliography{sd_ref}
	
\end{document}